\documentclass[11pt, a4paper]{article}
\pdfoutput=1

\usepackage[sc]{mathpazo} 
\usepackage[T1]{fontenc} 
\usepackage{microtype} 
\usepackage[hmarginratio=1:1,top=32mm,columnsep=20pt, left=15mm]{geometry} 
\usepackage{multicol} 
\usepackage[hang, small,labelfont=bf,up,textfont=it,up]{caption} 
\usepackage{booktabs} 
\usepackage{float} 
\usepackage{bookmark}
\usepackage{paralist} 
\usepackage{abstract} 
\usepackage{titlesec} 
\renewcommand\thesection{\Roman{section}} 
\titleformat{\section}[block]{\large\scshape\centering}{\thesection.}{1em}{} 
\titleformat{\subsection}[block]{\large}{\thesubsection.}{1em}{} 
\usepackage{amsmath,amssymb,latexsym} 
\usepackage{graphicx} 
\usepackage{array}
\usepackage{multirow}
\usepackage{csquotes}
\usepackage{tabularx}
\usepackage[usenames, dvipsnames]{color}
\usepackage{amsthm}
\usepackage{listings}

\usepackage{fancyhdr} 
\lstdefinelanguage{scala}{morekeywords={class,object,trait,extends,with,new,if,while,for,def,val,var,this},
otherkeywords={->,=>},
sensitive=true,
morecomment=[l]{//},
morecomment=[s]{/*}{*/},
morestring=[b]"}

\title{\vspace{-15mm}\fontsize{24pt}{10pt}\selectfont\textbf{A modular modelling framework for hypotheses testing in the simulation of urbanisation}} \author{
\large \textsc{Cl\'{e}mentine Cottineau$^{1, 2}$}\thanks{Corresponding author: \href{mailto:c.cottineau@ucl.ac.uk}{c.cottineau@ucl.ac.uk}},
\large \textsc{ Romain Reuillon$^{2}$}, \large \textsc{Paul Chapron$^{2, 3}$},\\
 \large \textsc{Sébastien Rey-Coyrehourcq$^{2}$,  
 \large \textsc{Denise Pumain$^{2, 4}$}}\\[2mm]
\small $^1$ Centre for Advanced Spatial Analysis, University College London, UK \\ 
\small $^2$ UMR 8504 G\'{e}ographie-cit\'{e}s and The Institute of Complex Systems Paris, France \\
\small $^3$ University of Lausanne, Switzerland, \small $^4$ Université Paris 1 Panthéon-Sorbonne, France \\
\normalsize  
\vspace{-5mm}
}
\date{}

\begin{document}

\maketitle 

\thispagestyle{fancy} 


\begin{abstract}

In this paper, we present a modelling experiment developed to study systems of cities and processes of urbanisation in large territories over long time spans. Building on geographical theories of urban evolution, we rely on agent-based models to 1/ formalise complementary and alternative hypotheses of urbanisation and 2/ explore their ability to simulate observed patterns in a virtual laboratory. The paper is therefore divided into two sections : an overview of the mechanisms implemented to represent competing hypotheses used to simulate urban evolution; and an evaluation of the resulting model structures in their ability to simulate - efficiently and parsimoniously - a system of cities (the Former Soviet Union) over several periods of time (before and after the crash of the USSR). We do so using a modular framework of model-building and evolutionary algorithms for the calibration of several model structures. This project aims at tackling equifinality in systems dynamics by confronting different mechanisms with similar evaluation criteria. It enables the identification of the best-performing models with respect to the chosen criteria by scanning automatically the parameter space along with the space of model structures.

\end{abstract}


\begin{multicols}{2} 

\section{Introduction}

Simulation models of urban systems were first developed in the 1950s and 1960s as a way to understand the complexity of cities and to forecast trends and consequences of planning policies. Several formalisms (thermodynamics, general systems theory, synergetic, microsimulation) were used \cite{batty2008}, following fashions as well as opportunities arisen from access to new technologies \cite{heppenstall2012, pumain2013}. Each of the methods cited have their specific advantages and drawbacks, that we will not discuss here, but all of them provide a same opportunity and challenge that is linked to simulation and has not changed as the formalisms evolved. The opportunity that we focus on in this paper is the function of {\bf virtual laboratory} that a computerised simulation model enables \cite{carley1999}, which is of paramount interest for human and social sciences in which {\it in vivo} experiments are impossible or difficult. By allowing to implement competing and/or complementary hypotheses into generative mechanisms\footnote{that is, a set of interaction activities performed by entities (cities) \cite{hedstrom2010} producing emergent patterns.} in a model testable against empirical data, this function of virtual laboratory makes the model a framework for the evaluation of the plausibility of different theories. However, simulation as a method and an epistemological way of testing theories gives way to a limitation known since Bertalanffy \cite{bertalanffy1968} as {\bf equifinality}. It describes the fact that even when a model is performing well, one cannot infer that the underlying combination of mechanisms is the one operating "in real life", because several models can lead to the same results (many-to-one), and the same process can lead to several patterns (one-to-many) \cite{osullivan2004}. The problem of adequacy assessment of the model with real life (also known as ontological adequacy testing, cf. \cite{rossiter2010}) is twofold. First, many effective (or "real") processes can lead to the same observed situation within the target system that we aim to model. Second, inadequate mechanisms implemented in the model can simulate in a satisfying way the situation under study. This causal challenge of identification of the generative mechanisms is a recurrent problem in social simulation for explanatory purposes \cite{gruneyanoff2009, elsenbroich2012}, but one that is usually overlooked at the stage of results' analysis. \\

We have tried in a previous study \cite{cherel2015} to tackle the one-to-many part of the challenge within the model, using an evolutionary algorithm to look for the maximum diversity in patterns produced by a given set of mechanisms in a model of systems of cities. What we present here relies to the multiplicity of possible causes leading to the single historical trajectory observed empirically. We present a multimodelling framework that allows to combine different mechanisms into a modular model evaluated against a unique set of evaluation criteria, enabling for the comparison of the performance of different model structures to simulate urbanisation and the evolution of a system of cities. \\

More precisely, we are interested in simulating the co-evolution of cities encompassed in a territorial system (typically a nation or a continent), and to reproduce the regular patterns or organised structures that are observed empirically in various systems of cities: their hierarchy, spacing and functional differentiation \cite{pumain1997}. Section \ref{sec:catalog} presents the patterns we aim to reproduce and the catalogue of theories and mechanisms on which we draw to compose the model. Section \ref{sec:multimodel} describes the multi-model and its implementation in our study case, as well as its exploration through multi-calibration. This exploration aims to explore the performance of different hypotheses in explaining the evolution of Soviet and post-Soviet cities. Section \ref{sec:conclu} concludes on the study case and the method.\\

\section{A catalogue of possible mechanisms of urbanisation}
\label{sec:catalog}

\begin{quote}
\small "It may also be useful to think of complex geographical models as extensions of thought experiments, where the necessary and contingent implications of theories can be examined. Further, admitting that 'all models are wrong' is akin to the realisation in post-structural social science that multiple competing accounts of the same settings are possible, and that faced with a diversity of accounts the context and intent of each must be an important element in the evaluation process." \cite[p.291]{osullivan2004}
\end{quote}

A large bunch of theories in geography, economics and natural sciences have tried to provide an account of the regularity of  urbanisation processes and the structuration of a system of cities, through models and narratives. Keeping in mind the equifinality challenge, this means that we already have a strong theoretical basis and several causal model candidates to confront with empirical regularities. Without being exhaustive on these theories, we try to provide an overview of the mechanisms that have been proposed in the literature. We then expose the kind of results that are achieved by statistical models using empirical data, and present the particularity of our study case.

\subsection{Competing theories}

Out of the three stylised facts describing the structure of a system of city: the hierarchy of cities by size, their regular spacing and the functional differentiation \cite{pumain1997}, the former has fostered the larger body of research, and will be our main criterion for stating the performance of a simulation model, whereas the latter two remain hard to formalise and to compare over time and space with respect to the availability of urban information at a local level. Consequently, we present in larger details the competing theories aiming at explaining the regularity of city size distribution and its evolution over time, and quickly review theories of location and functional specialisation.\\

The hierarchical organisation of city sizes represents a "mystery" \cite{krugman1996c} that has intrigued many researchers, because of its regularity and simplicity of description (the rank-size "rule") despite the complexity of urban functioning and interactions (for reviews, cf. \cite{cheshire1999, pumain2006}). \cite{auerbach1913, lotka1925, singer1936} are known to be the first to formalise this regularity, leading to the famous rule that the size of a city is a power law function of its rank in the hierarchy, a rule particularized by \cite{zipf1949} arguing that the Pareto exponent is expected to be -1. Alternative mathematical descriptions have been proposed, for instance the lognormal distribution \cite{gibrat1931, berry1961, eeckhout2004}. The two distributions however are close and tend to coincide in the middle-upper part of the hierarchy \cite{mitzenmacher2004, clauset2009}.\\

Because of this regularity, generative models of power laws and lognormal distributions have logically been considered as candidate explanations \cite{mitzenmacher2004, clauset2009}. The most famous one is Gibrat's law \cite{gibrat1931}, which generates a lognormal distribution by a process of "proportional effect". This model of multiplicative growth allocates randomly growth rates\footnote{If not stated otherwise, growth rates in this paper refer to the ratios of population variation during a time period by the total population at the beginning of the period.} from independent distributions, at each short time step, to cities independently from their size. It is thus considered close to a random walk. It results in amplifications of urban growth and decline leading to a lognormal distribution. The Simon model of preferential attachment \cite{simon1955} applied to cities generates power laws of city sizes by simulating an incremental creation of new urban blocks attached to existing urban clusters with a probability dependent on their size \cite{krugman1996a}. Those two models are elegantly simple enough to generate hierarchical distributions, but they lack an explaining power to help understand urbanisation processes: "Scaling laws often are viewed as over-identified: they can be generated by a wide range of distinct models. It is essential to select specifications that integrate in the model a significant part of the existing knowledge about towns and cities" \cite[p. 1]{pumain2004}. Some attempts were made to characterise them in terms of exogenous shocks and contingent local policies (for example in Gabaix's model \cite{gabaix1999}), but random growth models seem to miss the causal power of mechanism-based simulation models \cite{hedstrom2010}.\\

Other accounts of urban growth models leading to the observed stylized hierarchical structure and evolution (deterministic in Dimou and Schaffar's typology \cite{dimou2011}, by opposition to random models) range into two broad categories: equilibrium models of externalities (such as \cite{henderson1974, krugman1996b}) and evolutionary models of spatial interactions (such as \cite{allen1981, weidlich1983, bura1996}). The former type of models formalizes a balance of centripetal (sharing, matching, learning in the labour market for example) and centrifugal forces (pollution and congestion) resulting in optimal sizes for cities given the current technology. The latter type of models rely on attractivity and spatial interactions of cities to explain the dynamic of competition and cooperation resulting in a regular hierarchy in the urban system.\\

Location theories of cities typically build on two concepts to explain the regular spacing and empirical distribution of cities in space: site and/or situation advantages \cite{christaller1933, ullman1941, krugman1996b, pumain1997}. Site advantages refer to natural features available at an absolute location (natural resources, harbour conditions, etc.), whereas situation advantages refer to centrality in a transportation network or an interface relative position for example. All the theories mentioned above rely necessarily on relations with (consistent) territories providing resources or distance constraints for instance, in order to explain urban coevolution. \\
 
Theories of urban specialisation inherit from trade theories (via comparative and competitive advantages) and theories of product and innovation cycles. They state that cities have different features (size, situation, site, former specialisation) that make them akin to be more competitive to specialise in one economic sector in comparison with other cities, or to adopt an innovation sooner or later than their neighbours. Depending on the current cycle of the product, those specialisations affect not only the sectoral composition of the active population and its economic output, but also confers an advantage to early adopters and defines further options for specialisation and growth \cite{duranton2001, pumain2006al}.\\

Finally, political factors and singular policies are usually considered necessary to explain differentiated urban growth (through administrative functions or investment policies targeting specific cities at a given moment in time). \\
 
We compose our catalogue with mechanisms that formalize the theories cited above. They fall into five broad classes of generative mechanisms potentially accounting for the emergence of a structured system of cities: \\

  \textbullet{ { \bf Spatial Interactions} and diffusion allow for the exchange of information, money, goods and people. It thus makes cities co-evolve over time and adapt collectively to changing economic and innovation cycles through competition and cooperation, resulting in some complementarity of their specialisation. These local interactions and their consequences on the regular organisation of the system as a whole under spatial constraints could be thought of as "complex systems effects".} \\
   \textbullet{ { \bf Size effects} like agglomeration economies and urbanisation externalities illustrate a very direct and self-reinforcing cause for hierarchical differentiation. } \\
  \textbullet{ { \bf Site effects} explain the spatial location of growth processes around resource-rich areas for the related innovation cycle. } \\
    \textbullet{ { \bf Situation effects} illustrate the importance of the neighbouring relational environment (potential field, network accessibility, etc.) on a city's pattern of growth. } \\
   \textbullet{ { \bf Territorial effects} account for some exogenous (policy) shocks and the solidarity of urban trajectories in a common political space (through redistributive processes for example). } \\

We look for statistical evidences of these factors of urbanisation in the empirical literature before turning to our case study experiment, where we propose an implementation and evaluate the power of each of the five mechanisms into an agent-based (multi-) model.

\subsection{Empirical Results from the Literature}

\subparagraph{ Spatial Interactions} are tricky to measure because of the variety and non-commensurability of flows circulating between cities at various temporalities. Until recently, the diffusion of innovations (agricultural techniques \cite{hagerstrand1968}, telephone lines \cite{robson1973} or newspapers \cite{pred1973}) served as a proxy for these interactions. Since the development of various volumes of high velocity data, actual interactions (like phone calls \cite{krings2003}) have confirmed for example the relevance of the gravity model to describe inter-city interactions.

\subparagraph {Size effects} in urban growth and differentiation were revealed by a persistent empirical correlation between growth rates and city sizes over long periods of time. All over the 19th century, \cite[p. 79]{robson1973} measured a positive coefficient between the log of English and Welsh cities' population and their ten-year growth rates (from a minimum of +1.47 between 1861 and 1871 to a maximum of +8.53 between 1821 and 1831). This correlation is found for French cities as well \cite{guerinpace1995}. The size effect finally relates to the stability of the rank position of large cities (by comparison with the fluctuations of smaller cities).
   
\subparagraph{ Site effects} were classically approached by estimating the surplus of growth associated with a localised resource (typically, deposits of natural materials, such as coal or gas). In the Soviet urban system of the 1920s-1930s, the location of a city on a coal deposit was associated with a surplus of 1.15 points in percentage of average demographic growth per annum, everything else being equal. A surplus over 0.5 point is observed nowadays (1989-2010) for oil and gas deposits \cite{cottineau2014}. In the USA, \cite{black2003} estimate that the coastal location (e.g. a resource for tourism) is associated with a significantly higher ten-year growth rates of 3 to 5 points.
  
\subparagraph{ Situation effects} can be revealed by the spatial autocorrelation of growth or the coevolution between transportation networks and urban networks. In the first case, \cite{hernando2015} found a characteristic distance for spatial autocorrelation of growth rates of 215 km for American counties and of 80 km for Spanish cities. As for transportation dynamics, \cite{bretagnolle2003} measured the correlation between accessibility and growth rates for French cities in the last two centuries. She finds that cities that were weakly connected (by any transportation network: road, rail or air) in 1900 and stayed isolated in 2002 grew slower (0.94\% on average per annum between 1900 and 2002) than cities that became motorway nodes (1.21\%) or multimodal hubs (1.69\%). Likewise, well connected cities at the beginning of the period tended to grow faster than the first category.
    
\subparagraph{ Territorial effects} can be approached empirically by relating political statuses to dynamics of growth. In developing countries, regional capitals were found to grow significantly faster by 0.5 to 1 point of annual average growth rate in the 1960s \cite{preston1979} and the 1990s \cite{brockerhoff1999}. In the Former Soviet Union, the regional status of capital has proven important to predict urban growth \cite{harris1970}, the coefficient regressed against growth rates over time ranges from +0.24 point between 1989 and 2002 to +1.88 between 1926 and 1939 \cite{cottineau2014}. Besides, cities that belong to the same territory have shown an increased pattern of synchronicity in their growth and decline trajectories from the 1980s on, suggesting evidence of both political shocks and territorial solidarity in the spatial distribution of urban growth.\\

Despite the corroboration of theories provided by the numbers cited above, statistical correlations do not prove any causal chain and fail to explain the processes at work. Moreover, they are not adapted to model dynamic, non-linear and complex evolutions and interactions. This is a recurrent problem that has led social scientists to promote a combination of statistical methods with generative (simulation) modelling \cite{byrne1998, goldthorpe2001}. 

\subsection{A Case Study : the Former Soviet Union}
\label{sec:eval}

We applied such a mixed methodology to the case study of the urban Soviet Union of the last 50 years \cite{cottineau2014, cottineau2015}. This system is mainly interesting to us because it has the reputation of having been a controlled economic, political and social system aiming at reaching equalisation (of city size for example) and yet, while observing its evolution with generic urban models, the urban trends appear very classic. By this we mean for instance that the evolution of the percentage of urban population follows a very classical logistic function of urban transition, that the rank-size slope is close to the value expected with respect to the time of settlement of the different parts of the territory, and that it has increased over time, indicating an increase of inequality of city sizes (despite anti-urban political discourses \cite{cottineau2014}). The monographic explanation of urban growth in the Former Soviet Union has been dominant in urban geography, and we argue that it might have hidden very generic processes of urbanisation in this system of cities \cite{cottineau2016}. \\

Our modelling experiment aims at testing the genericity of Soviet and post-Soviet urbanisation by simulating generic mechanisms and testing them against harmonised historical data. The database used in this study consists of open demographic, territorial and site informations on 1929 agglomerations of the Former Soviet Union between 1840 and 2010 \cite{darius2014}. \\

We analyse and compare the explaining power of generic mechanisms in a simulation of the period before (1959-1989) and the period after (1989-2010) the dislocation of the Soviet Union. The agent-based model framework of this experiment is called MARIUS\footnote{Models of Agglomerations in the perimeter of Imperial Russia and the former Soviet Union. For further details, see \cite{cottineau2015}}. In this model, cities interact as collective agents \cite{bura1996} and we evaluate every simulations with three evaluation criteria. The first two criteria are boolean indicators of the "realism" of the microdynamics of the simulation : we only analyse simulations in which the number of cities with a nil wealth and the number of cities producing more output in one year than their stock of wealth is 0. Once these controlling criteria are met, we try to minimise the distance $\delta$ between the simulated population $P_s$ and the observed population $P_o$ of each city $i$ at each date $t$ we have census information for, as stated in equation \eqref{eq:eval}. 
\begin{align}
\label{eq:eval}
\delta = \sum_{t} (\sum_{i} {(log(P_{o,i,t}) - log(P_{s,i,t})})^2)
\end{align}

This criterion ensures that a "perfect" model (achieved for $\delta = 0$) would predict the right amount of urban growth and its exact location over time. In order to compare different time periods, we normalise $\delta$ by the number of cities and the number of censuses used in a given period.

\section{Modular multimodelling experiment}
\label{sec:multimodel}

The incentive to implement competing and complementary theories into different models evaluated against one another is a recurrent plea in the simulation literature \cite{openshaw1983, osullivan2004, batty2005, thiele2015, batty2015}. It reveals how tricky its implementation and automatic evaluation might be, besides the epistemological challenge of equifinality and the kind of conclusions one can draw from this confrontation. Indeed, thirty years after the "Automated Modeling System to Explore a Universe of Spatial Interaction Models" by \cite{openshaw1983}, there are no standard tools nor formal methodology for theory testing with simulation models. Indeed, Openshaw's automated way to explore model structures, being a pure optimisation way of discovering model structures, is impressive methodologically but it does not suit our goal of theory testing in a virtual laboratory, because it can result in optimal models that are impossible to interpret. \\

Instead, we think that the first step should be to gather a catalogue of theoretical processes and mechanistic hypotheses working as potential explanations. This usually precedes the mixed-modelling step \cite{contractor2000, auchincloss2008} and prevents from endlessly building models "from scratch" \cite{thiele2015}; it allows to build on previous work and experience.\\

In a second step, we have built a baseline model of urban interactions and implemented hypotheses as blocks of mechanisms that can be activated or discarded to compose a family of simulation models (section \ref{sec:mechanisms}). By automatically calibrating all model structures against a similar evaluation objective (section \ref{sec:multicalibration}), we were able to compare the power of several groups of hypotheses in the case of the Soviet and post-Soviet urbanisation (section \ref{sec:results}).

\subsection{Implementing mechanisms as building blocks}
\label{sec:mechanisms}

The multi-model is composed of the baseline model and modules of mechanisms that override the sequence of agents' rules when they are activated. In the following sections, we detail the baseline model and the modular blocks of mechanisms that are added incrementally to the original equations. For further informations, the baseline model and the first two additional mechanisms were described and evaluated in details in \cite{cottineau2015}.

\subsubsection{The Baseline Model}

The baseline model relies on the assumption that population and wealth are the basic descriptors of cities and the engine of their coevolution. It therefore models exclusively size effects of population on wealth and spatial interactions.\\

At initialization, each city $i$ of the Former Soviet Union is setup with its historical population $P_i$ at the beginning date of simulation, and located at its empirical coordinates, enabling site, situation and distance to play in the same geometry as in the target system. An estimated value of wealth $W_i$ (expressed in a fictive unit) is determined for each city with respect to its size\footnote{This first mechanism is a first possibility of implementing the theoretical hypothesis of agglomeration economies. Wealth is indeed distributed superlinearly for each value of $populationToWealth$ significantly greater than 1.}, following equation \eqref{eq:init}.\\

\begin{align}
\label{eq:init}
\small W_i &= P_i^{populationToWealth} \\
\small & populationToWealth \ge 1
\end{align}

Time is modelled as discrete steps, each of which represents a time period of one year, during which interactions occur in a synchronous way. At each step: \\

\subparagraph{ Each city $i$ updates its economic variables:} a global supply $S_i$ \eqref{eq:supply} and a global demand $D_i$ \eqref{eq:demand}, according to its population $P_i$ and three parameters ($economicMultiplier$, $sizeEffectOnSupply$ and $sizeEffectOnDemand$).

\begin{align}
\label{eq:supply}
\small S_i &= economicMultiplier \times P_i^{sizeEffectOnSupply} \\
\small & economicMultiplier > 0\\
\small & sizeEffectOnSupply \ge 1
\end{align}

\begin{align}
\label{eq:demand}
\small D_i &= economicMultiplier \times P_i^{sizeEffectOnDemand} \\
\small & sizeEffectOnDemand \ge 1
\end{align}

\subparagraph{ Each city interacts with other cities} according to the intensity of their potential $IP$ \eqref{eq:potential}. For two distinct cities $i$ and $j$, the computation of the interaction potential $IP_{ij}$ consists in confronting the supply of $i$ \eqref{eq:sij} to the demand of $j$ with an equation borrowed to the gravity model \eqref{eq:dji}. 

\begin{align}
\label{eq:potential}
IP_{ij} &= \frac{S_i \times D_j}{d_{ij}^{distanceDecay}}\\
& distanceDecay \ge 0
\end{align}
{\small with $d_{ij}$ a measure of distance between $i$ and $j$.\\}

Interactions of cities $i$ and $j$ based on their potential $IP_{ij}$ result in a transaction $T_{ij}$ from $i$ to $j$ \eqref{eq:transaction}.

\begin{align}
\label{eq:sij}
S_{ij} = S_i \times \frac{IP_{ij}}{\sum_{k} {IP_{ik}}}
\end{align}

\begin{align}
\label{eq:dji}
D_{ij} = D_i \times \frac{IP_{ji}}{\sum_{k} {IP_{ki}}}
\end{align}

\begin{align}
\label{eq:transaction}
T_{ij} = \min[S_{ij}, D_{ji}]
\end{align}

\subparagraph{ Each city updates its wealth $W_i$} according to the results of its transactions $T$ (unsold supply $US_i$ \eqref{eq:unsold} and unsatisfied demand $UD_i$ \eqref{eq:unsatisfied}) in which it was committed \eqref{eq:wealth}.

\begin{align}
\label{eq:wealth}
W_{i,t} = W_{i,t-1} + S_i - D_i - US_i + UD_i
\end{align}

\begin{align}
\label{eq:unsold}
US_i = S_i - {\sum_{j} {T_{ij}}}
\end{align}

\begin{align}
\label{eq:unsatisfied}
UD_i = D_i - {\sum_{j} {T_{ji}}}
\end{align}

\subparagraph{ A simulation step ends when each city updates its population} according to its new resulting wealth \eqref{eq:updatepop} : 
\begin{align}
\label{eq:updatepop}
P_{i,t} = P_{i,t-1} + \frac{W_{i,t}^{wealthToPopulation}-W_{i,t-1}^{wealthToPopulation}}{economicMultiplier}
\end{align}

\subsubsection{Mechanism Increments}

	\subparagraph{The mechanism that accounts best for interactions benefits} at the intercity level is the one called {\bf bonus}. It
\begin{quotation}
"[...] features a non-zero sum game [...], rewarding cities who effectively interact with others rather than internally. We assume that the exchange of any unit of value is more profitable when it is done with another city, because of the potential spillovers of technology and information." \cite{cottineau2015}
\end{quotation}

This bonus $B_i$ depends on the volume of transactions and the diversity of partners $J_i$ the city $i$ has exchanged with \eqref{eq:bonus}. It is added to the wealth at the end of each step \eqref{eq:wealthbonus}, following equation \eqref{eq:wealth}:

\begin{align}
\label{eq:bonus}
B_i = bonusMultiplier \times \frac{(\sum_{j} {T_{ij}} + \sum_{j} {T_{ji}}) \times J_i}{n} 
\end{align}
{\small $n$ being the total number of cities in the system (i.e. 1145 for a simulation beginning in 1959, 1822  for a simulation beginning in 1989), and $J_i$ the number of partners with which $i$ has engaged in the current simulation step.}

\begin{align}
\label{eq:wealthbonus}
W_{i,t} = W_{i,t} + B_i
\end{align}

\subparagraph{A mechanism related to situation advantages} is called {\bf fixed costs}. It ensures that the situation of each city in the system is taken into account in its interactions with other cities. 

\begin{quotation}
"Every interurban exchange generates a fixed cost (the value of which is described by the free parameter $fixedCost$). This implies two features that make the model more realistic: first, no exchange will take place between two cities if the potential transacted value is under a certain threshold ; second, cities will select only profitable partners and not exchange with every other cities. This mechanism plays the role of a condition before the exchange." \cite{cottineau2015}
\end{quotation}

The interaction potential between city $i$ and city $j$ will be positive only if the potential value that $i$ is willing to sell to $j$ is superior to the fixed value it costs it to negotiate, prepare and send the transaction \eqref{eq:fixedcosts}:
\begin{align}
\label{eq:fixedcosts}
IP_{ij}  = \begin{cases}
  IP_{ij} & \text{if } S_{ij} > fixedCost, \\
  0 & \text{otherwise}.
\end{cases}
\end{align}

Therefore, each transaction of a city $i$ gives way to a fixed cost. Their sum is subtracted from the wealth of city $i$ at the end of each step \eqref{eq:wealthfixedcost}, following equation \eqref{eq:wealth}:

\begin{align}
\label{eq:wealthfixedcost}
W_{i,t} = W_{i,t} - J_i \times fixedCost
\end{align}

\subparagraph{Site effects} are targeted by the {\bf resource} mechanism: site advantages are particularised in this model by natural resource deposits (more specifically: coal deposits $C$ on the one hand, and oil and gas deposits $O$ on the other hand). The assumption is made that if the city $i$ is located on some coal or oil deposits ($C_i = 1$ or $O_i = 1$), the city benefits from the advantage granted by the extraction activity. The capacity of extraction depends on the capital (wealth) of the city and takes the form of a wealth multiplier for each resource \eqref{eq:resource} after equation \eqref{eq:wealth}:
 \begin{align}
\label{eq:resource}
W_{i,t}  = W_{i,t} \times (1 + \begin{cases}
  coalEffect & \text{if } C_i = 1, \\
  oilAndGasEffect & \text{if } O_i = 1
\end{cases})
\end{align}

\subparagraph{Territorial and political effects} are formalised by the {\bf redistribution} mechanism. It allows for a redistribution of wealth between cities of the same territory $R$ (region or State). To do so, territorial taxes $tt_k$ are collected in each city $k_R$, as a proportion $territorialTaxes$ of their wealth. The total amount of taxes collected is $TT_R$\eqref{eq:taxes}:

\begin{align}
\label{eq:taxes}
TT_{R} = \sum_{i} {tt_{i,R}} = \sum_{i} {territorialTaxes \times W_{i,R}}
\end{align}

From this taxes, the administrative status of the territory $R$ (denoted by $CC_{i,R}$, set to 1 if $i$ is the capital city of the region and 0 otherwise) allows the capital city to take a share $CS$ for its administration needs \eqref{eq:capital}:

\begin{align}
\label{eq:capital}
CS_{R} = capitalShareOfTaxes \times TT_{R}
\end{align}

The rest of the taxes is redistributed to cities of region. Each city $i_R$ receives a share $tr_{i,R} $ that is proportionate to its population \eqref{eq:redistribution}:

\begin{align}
\label{eq:redistribution}
tr_{i,R} = (TT_{R} - CS_{R}) \times \frac{P_{i,R}}{\sum_{k}{P_{k,R}}}
\end{align}

The balance of the territorial redistribution is added to the wealth of a city \eqref{eq:wealthredistribution} after equation \eqref{eq:wealth}:
\begin{align}
\label{eq:wealthredistribution}
W_{i,t} = W_{i,t} - tt_{i,R} + tr_{i,R} + \begin{cases}
  CS_{R} & \text{if } CC_{i,R} = 1, \\
  0 & \text{otherwise } 
\end{cases}
\end{align}

\subparagraph{Finally, territorial and situation explanations} are mixed in the {\bf urban transition} mechanism. To account for the different opportunities of cities to attract rural migrants in the different regions, we model the evolution of the urban transition curves over time. As shown empirically  \cite{cottineau2014}, 100 out of the 108 regions of the Former Soviet Union have followed the scheme of the urban transition. It means that their urbanisation rate $U_R$ (in \%) has followed a logistic function over time $t$ \eqref{eq:urbantransition}:
 \begin{align}
\label{eq:urbantransition}
U_{R,t} = \frac{100}{1 + e^{- urbanisationSpeed \times t}}
\end{align}

We have estimated the parameter $urbanisationSpeed$ on empirical data and normalised the time units in order to position every region in a single urbanisation curve with respect to their time lags based on their urbanisation rate. The consequence of this initialisation process is that each region will move one step further on the urbanisation curve (leading eventually to 100\%) at each simulation step, but that the rural potential to migrate in cities will depend on its current position on the urban transition curve (high potential for weakly urbanised regions, small potential for regions already very urban). The migration potential of each city $i$ in territory $R$ is built as a multiplier $TM_R$ specific to each region for each time step \eqref{eq:territorialMultiplier}:
 \begin{align}
\label{eq:territorialMultiplier}
TM_{R,t} = 1 + (1 - U_{R,t} \times ruralMultiplier)
\end{align}

This extra population growth is added \eqref{eq:ruralpop} after \eqref{eq:updatepop} and the new urbanisation rates of regions are updated for $t+1$ \eqref{eq:urbantransition}:
  \begin{align}
\label{eq:ruralpop}
P_{i,R,t} = P_{i,R,t} \times (1 + TM_{R,t})
\end{align}

Because we want to evaluate the contribution of each theoretical mechanism to the simulation of urbanisation and its itneraction with other mechanisms, we need a model that enables modules to be activated and de-activated. The technical implementation of MARIUS permits the modular functioning of the model, i.e. with or without any of its building blocks.

\subsubsection{Technical modular implementation}

In order to achieve a modular implementation of the model and its mechanisms, we leveraged the mixin-methods system of the Scala programming language \cite{scharli_traits:_2003}. The mixins were first proposed at the beginning of the 90s in the Jigsaw programming language \cite{bracha_programming_1992}. They are now adopted in mainstream languages such as Scala. It has been established as a powerful way of to perform type-safe dependency injection framework \cite{wampler2009programming}, which is a powerful paradigm to achieve modularity. The mixin pattern allows to achieve type safe dependency injection. Mixins are therefore a suited tool to achieve modularity in model implementations, which we use to implement MARIUS in Scala language\footnote{ The source code of the model is available under open-licensing here: \url{https://github.com/ISCPIF/simpuzzle}}. \\

The mixin-methods concept \cite{mixins, lucas1994modular} generalizes object-oriented programming to enable feature-oriented programming \cite{prehofer_feature-oriented_1997}. It makes the definition of class hierarchy more flexible \cite{Steyaert1995}. Variations of isolated features (such as the different update functions of city wealth and population in our model) are defined in separate modules and mixed-in with each other at the object instantiation point, also called \textit{mixin application}. The advantage of mixins (or trait) over classical object-oriented programming is the possibility of defining numerous variations of several features without increasing exponentially the number of specialized class.\\

For instance, let's consider a class \textit{C} implementing two methods \textbf{a} and \textbf{b}\footnote{ We could think of class \textit{C} as the interaction potential function, \textbf{a} representing the basic implementation of equation \ref{eq:potential} and \textbf{b} the fixed cost selection of equation \ref{eq:fixedcosts}.}. To define alternative implementations of those methods in the classical object-oriented paradigm, one would implement subclasses of \textit{C} and override the implementations of \textbf{a} and \textbf{b} in each subclass. For instance in the listing \ref{lst:object} the class \textit{C1} specialises \textit{C} and defines implementations for the methods \textbf{a} and \textbf{b}.

\begin{lstlisting}[caption={Object oriented specialisation},label={lst:object}]
abstract class C extends A with B {
  def a(x: Double): Double
  def b(x: Double): Double
  def c(x: Double) = /* Compute something using a and b */
}

class C1 extends C {
  def a(x: Double) = /* Some implementation of a */
  def b(x: Double) = /* Some implementation of b */
}
\end{lstlisting}

This pattern achieves a very low level of code reusability. Let \textbf{a} and \textbf{b} have 10 possible implementations, then 100 specialised implementations of \textit{C} would be required. The mixins method solves the problems of combinatorial explosion of the number of implementations by delaying the entanglement of the class components at the instantiation site. Listing \ref{lst:mixin} exposes an implementation based on mixins providing alternative implementations of \textbf{A} and \textbf{B} and the corresponding parameter specifications. The implementation choice is delayed until the instantiation point (last lines of listing \ref{lst:mixin}) at which a mixin is defined. 

\begin{lstlisting}[caption={Mixin in Scala},label={lst:mixin}]
trait A {
  def a(x: Double): Double
}

trait B {
  def b(x: Double): Double
}

trait C extends A with B {
  def c(x: Double) = /* Compute something using a and b */
}

// Implementation 1 of trait A
trait A1 extends A {
  def a(x: Double) = /* Some implementation */
}

// Implementation 2 of trait A
trait A2 extends A {
  // Parameter p0 used in this version of a
  def p0: Double
  def a(x: Double) = /* Some implementation using p0 */
}

// Implementation 1 of trait B
trait B1 extends B {
  def b(x: Double) = /* Some implementation */
}

// Implementation 2 of trait B
trait B2 extends B {
  def b(x: Double) = /* Some implementation */
}

val instance1 = 
  new C with A1 with B2 {}

val instance2 = 
  new C with A2 with B2 {
    // Value for parameter p0 of trait A2
    def p0 = 1.0
  }

\end{lstlisting}

Scala traits expressiveness can be leveraged to implement modular evolutive type-safe modelling frameworks proposing alternative model features, feature composability and the formalisation of the feature dependencies. The implementation of alternative behaviours in several traits provides the isolation of model component implementations and explicit dependencies between these components. Each component defines free parameters that have to be set at the model instantiation site, otherwise it won't compile.\\

In this experiment we defined each alternative model mechanism in a particular trait. The executable model has then be composed by picking the traits we wanted to test. In order to evaluate concurrently all the alternative mechanisms, we generated all the possible models (or combination of mechanisms). It was achieved using a code generation algorithm which produces all the possible models implementations by generating a Scala source code containing all the possible traits combinations, such as the one shown on Listing \ref{lst:generated}.

\begin{lstlisting}[caption={Example of generated code},label={lst:generated}]

def model(index: Int, parameters: Seq[Double]) = 
  index match {
    case 0 => 
      new Model with T11 with T21 with ... {
        def p0 = parameters(0)
        def p1 = parameters(1)
        ...
      }
    case 1 =>
      new Model with T11 with T22 with ... {
        def p0 = parameters(0)
        def p1 = parameters(1)
        ...
      }
    case 2 => ...
  }


\end{lstlisting}

This generated source code implements a single function encapsulating all the model alternatives. This function takes two arguments: an index of the implementation that shall be executed and a vector of parameters to set for the model (for this work we calibrated only double precision floating points values). Note that the vector has a fixed size which does not depends on the model instantiated. A given model implementation generally does not use all the parameters: instead it will use only some of them and ignore the others. 

\subsection{Calibrating a multi-model}
\label{sec:multicalibration}

\begin{quote}
\small "Whatever changes occur in the institutional, political and social context of computational models, the question of how to learn from models remains. It is clear that assessment of the accuracy of a model as a representation must rest on argument about how competing theories are represented in its workings, with calibration and fitting procedures acting as a check on reasoning." \cite[p.291]{osullivan2004} 
\end{quote}

In order to evaluate the capacity of the models to reproduce the historical trajectory of urbanisation in the Former Soviet Union, we rely on an automated calibration. This procedure is part of the model evaluation \cite{grimm2012, thiele2014, schmitt2015}. Its aim is to find the values of parameters for which the model results match the fitness criteria (here: $\delta$, the lowest possible distance to the data, the realistic criteria having been met). If the model can be calibrated, then it is shown that the mechanisms included in the model are sufficient to simulate the urban trajectory (not necessary though). If there are no parameter value for which the fitness criteria is met, then the combination of mechanisms is not sufficient to model urbanisation under the current implementation.\\

In order to calibrate all the models at once, we designed a variant of the NSGAII genetic algorithm that includes a niching mechanism \cite{mahfoud_niching_1995}. Niching methods aim at preserving suboptimal solutions to preserve diversity. Our niching algorithm divided the population into sub-populations, each sub-population containing one model alternative. The genome of each individual\footnote{An individual correspond to a vector of parameter values and structure index that defines the model under evaluation} contains two parts. The first part is an integer value that corresponds to the index of the model alternative on which the genome is evaluated. The second part is a vector of double values containing the values of all the parameters for the model. \\

In order to evaluate a genome, we designed a fitness function. This function calls the generated function described above, runs the model and evaluates its dynamics using the fitness function described in section \ref{sec:eval} (i.e. the criteria of realism of micro-dynamics and the distance between simulated and observed population data for each city at each census date).\\

In NGSAII, the elitism operation preserves the best individuals among the whole population. The evaluation algorithm we applied has the exact same elitism strategy for each sub-population. No global elitism strategy was performed, instead we kept the 50 best-performing individuals in each sub-population (or model combination of mechanisms). In order to speed up the convergence of the algorithm, we also tweaked the mutation operation: it had a $10\%$ chance of mutating the "model index" part of the genome. The new value for the "model index" was drawn uniformly among the possible model indexes. This allowed to periodically test on other models, some parameter values that were performing well on a given model.\\

We then distributed this algorithm on the European grid EGI using the technique known as the "island model" in the same way as we described in \cite{schmitt2015}. We ran 200 000 jobs of 2 hours.

\subsection{Hypothesis testing to explain urbanisation in the Former Soviet Union}
\label{sec:results}

After the multicalibration procedure, we end up with the best performing simulations of 64 different structures of models\footnote{For example: baseline model, baseline model + fixed cost, baseline model + urban transition + resources, baseline model + redistribution + urban transition, baseline model + fixed cost + bonus + resources, etc.}. The analysis of calibration results consists in relating the performance of these calibrated models (in terms of distance between simulated and observed growth hierarchically and spatially) to their structure and the values of parameters associated with each activated mechanism. This analysis was made using an interactive application available online: \url{http://shiny.parisgeo.cnrs.fr/VARIUS}.\\

We present the results of this exploration in the form of three questions at the macro, meso and micro scale of the city-system. 

 \subparagraph{1/ Which is the most parsimonious model to simulate the evolution of cities before and after the collapse of the Soviet Union?} For simulations starting in 1959 up to 1989\footnote{The Soviet Union actually came to an end in Decembre 1991. In this study, we rely on historical data and therefore consider the last Census date of the Soviet Union in 1989 as the transition point between the two economic, political and territorial regimes.}, the best model with only one additional mechanism is composed of the baseline model plus the mechanism of urban transition (cf. Table \ref{tab:1mech1959}).\\

\begin{figure*}
 \begin{center}
  \caption{Residuals of the parsimonious models}
  \label{fig:res1mech}
              Urban Transition model. Simulation 1959-1989         |            Resource Extraction model. Simulation 1989-2010
\includegraphics[width=0.45\textwidth]{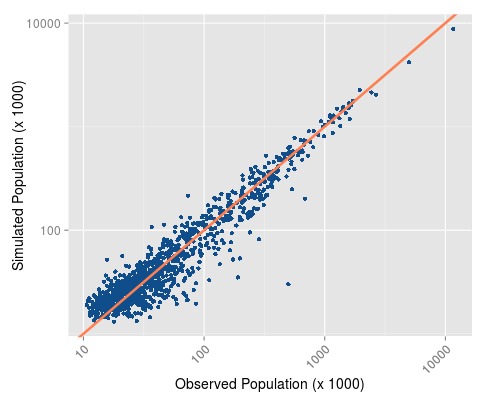}   
\includegraphics[width=0.45\textwidth]{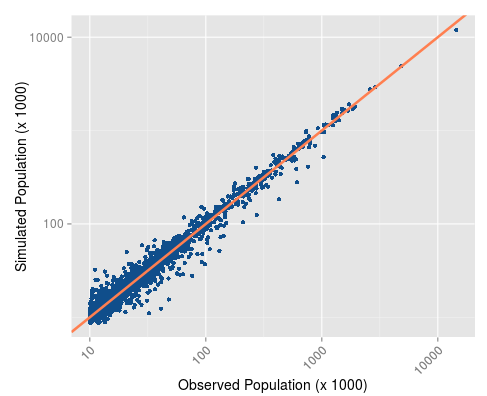}   \\ 
Urban Transition model 1959-1989 ($| residuals | \ge 0.3$)
  \includegraphics[width=0.8\textwidth]{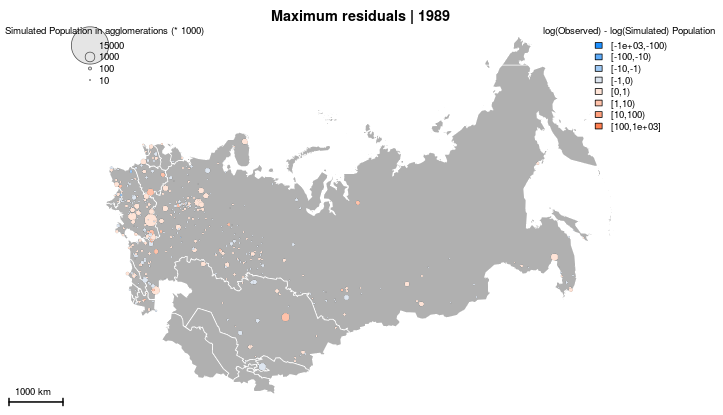}   \\ 
  Resource Extraction model 1989-2010 ($| residuals | \ge 0.3$)
  \includegraphics[width=0.8\textwidth]{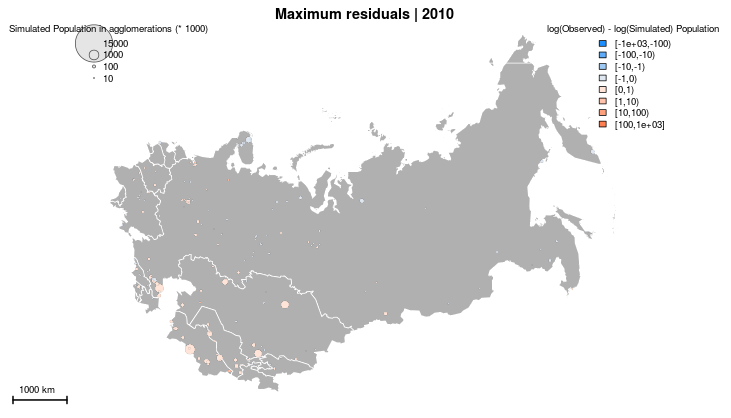}   
 \end{center}
 \end{figure*}

It is characterised by significant economies of agglomeration ($sizeEffectOnSupply = 1.05$) but a linear function of demand with size. The rural multiplier is equal to 3.5\% and allows to simulate fast urbanising regions of Siberia and Central Asia (cf. figure \ref{fig:res1mech}). However, the population of a majority of cities is under-estimated in the simulation, especially in the upper part of the hierarchy, around Moscow and in eastern Ukraine\footnote{or more generally in the Western part of the Former Soviet Union. See question 3 for an hypothesis as to why that might be.}.

\begin{table}[H]
\centering
\small
\caption{Best simple model before the transition}
\label{tab:1mech1959}
\begin{tabular}{|c|c|c|}
\hline
Parameter name & Value & Mechanism \\ \hline
\small economicMultiplier &  \small 0.002193758 & \small Generic  \\
\small populationToWealth & \small 1.000184755 & \small Generic  \\
\small sizeEffectOnSupply & \small 1.053943022 & \small Generic  \\
\small sizeEffectOnDemand &\small  1.000000000 & \small Generic  \\
\small wealthToPopulation &\small  0.203567639 & \small Generic  \\
\small distanceDecay & \small 1.872702086 & \small Generic  \\
\small ruralMultiplier & \small 0.034975771 & \small UrbanTransition  \\  \hline
Normalized $\delta$  & n cities & Time steps \\ \hline
\small 0.01423387 & \small 1145 & \small 30  \\ \hline
\end{tabular}
\end{table}

After the transition, the best performing model with one additional mechanism includes site advantages for cities. This model has two additional parameters and meets the evaluation criteria (per city per census) twice better than the best model for the previous period (0.005 vs. 0.01, cf. tables \ref{tab:1mech1959} and \ref{tab:1mech1989}). \\

The analysis of the parameters fits the empirical observations of faster growing cities located on oil and gas deposits ($oilAndGasEffect = 0.02$), and declining cities in the Donbass and Kuzbass coal regions ($coalEffect = -0.01$). This model's specifications include very low size effects and a very uneven wealth distribution amongst cities at initialisation ($populationToWealth = 1.12$). The residuals are distributed roughly symmetrically (figure \ref{fig:res1mech}), but without any mechanism of urban transition, the model clearly underestimates the post-1989 growth of all the rapidly growing cities of Central Asia. \\

\begin{table}[H]
\centering
\small
\caption{Best simple model after the transition}
\label{tab:1mech1989}
\begin{tabular}{|c|c|c|}
\hline
Parameter name & Value & Mechanism \\ \hline
\small economicMultiplier &  \small 0.502616330 & \small Generic  \\
\small populationToWealth & \small 	1.124963276 & \small Generic  \\
\small sizeEffectOnSupply & \small 1.002982515 & \small Generic  \\
\small sizeEffectOnDemand &\small  1.000808442 & \small Generic  \\
\small wealthToPopulation &\small  0.699943763 & \small Generic  \\
\small distanceDecay & \small 1.475836151 & \small Generic  \\
\small oilAndGazEffect & \small 0.017066495 & \small Resources  \\ 
\small coalEffect & \small -0.011792670 & \small Resources  \\  \hline
Normalized $\delta$ & n cities & Time steps \\ \hline
\small 0.005180008 & \small 1822 & \small 21  \\ \hline
\end{tabular}
\end{table}

To summarise, situation effects and territorial effects seem to be the dominant candidates for explaining the specific part of the trajectories of cities in the Soviet era, while site effects seem to have taken over since 1991, the transition to capitalism and the rise of oil prices in the world markets. Moreover, the better fit of the latter model could be an indication of the "normalization" of the urban processes in the post-Soviet space, compared to a more singular pattern of Soviet urbanisation.

 \subparagraph{2/ Which are the mechanisms (and mechanisms' interactions) that are essential to model the Soviet and post-Soviet urbanisation patterns?}
 
 We statistically analyse the results of the multicalibration to evaluate the contribution of each mechanism (everything else being equal in the model structure) to the reduction of distance between simulated and observed demographic data for each city. 

 \begin{figure}[H]
 \begin{center}
  \caption{Contribution of mechanisms to a simulation that reduces the distance to observed data}
  \label{fig:contribRes}
  \includegraphics[width=0.45\textwidth]{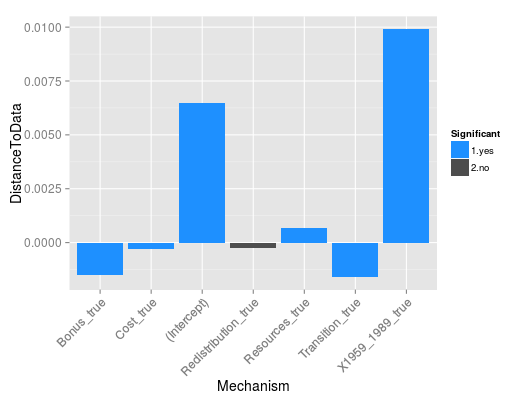}   
 \small N.B.: Estimated coefficients are considered significant for p-values inferior to 0.005
 \end{center}
 \end{figure}

We find that on average, the bonus, fixed costs and urban transition mechanisms tend to reduce the simulation error significantly (cf. figure \ref{fig:contribRes}). The transition and bonus mechanisms are identified to be the most effective ones. By contrast, the resource mechanism is correlated significantly with a change in the evaluation criteria, but tends to increase the error when it is activated (compared with model structures without this mechanism). This counter-intuitive result might be linked to the weak influence of resource extraction for the first period of simulation, when there is a diversity of urban trajectories within resource-rich regions. The redistribution mechanisms does not appear significant on average in this analysis, as it plays a minor role in the reduction of error when activated. \\

Finally, we confirm the observation that our models work better to simulate the urban evolution following the dismantlement of the Soviet Union. Indeed, as shown in figure \ref{fig:contribRes}, the value of normalized $\delta$ is greater by 0.01 point on average when the model is specified for the first period. This could be an expression of "normalisation" or simplification of the urban processes in the post-Soviet space after the political and economic transition of the 1990s.

 \subparagraph{3/ What are the cities that resist modelling? In other words, what are the cities that are too specific to be modelled by any of the mechanisms implemented?} To answer this last question, we statistically analyse the difference between the log of the population observed and the log of the population simulated at the last evaluation Census date for cities included in the simulation, with respect to their locational and functional attributes. The models for which we present the results below contains all the implemented mechanisms, and are applied to the two periods of enquiry. \\

\begin{figure*}
 \begin{center}
  \caption{Profiles of residuals}
  \label{fig:resProfiles}
Complete model. Simulation 1959-1989         |           Complete model. Simulation 1989-2010
\includegraphics[width=0.45\textwidth]{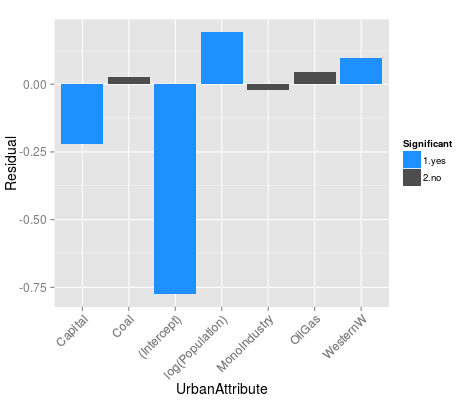}   
\includegraphics[width=0.45\textwidth]{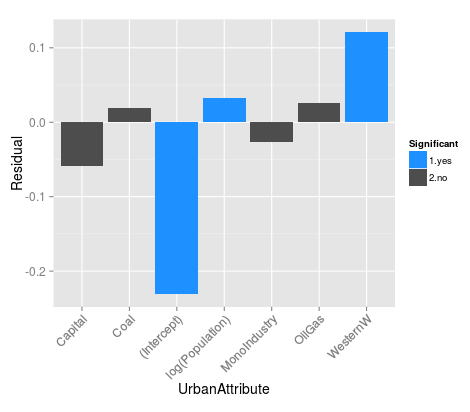}   \\ 
  \small N.B.: Estimated coefficients are considered significant for p-values inferior to 0.01
 \end{center}
 \end{figure*}

For the two periods, we find that our models persistently and significantly under-estimate the growth of the largest and most western cities of the (Former) Soviet Union, everything else being equal (cf. figure \ref{fig:resProfiles}). Moreover, capital cities appear to have grown less historically than what we can predict with a complete model of the period 1959-1989. The other urban attributes included in the regressions (natural resources and mono-functional specialisation) do not seem associated with any systematic pattern of over- or under-estimation.\\

The difficulty to reproduce the trajectory of the largest cities has been encountered for a comparable model of system of cities (Simpop2, see \cite{bretagnolle2010}) and solved by the exogenous introduction of innovations to account for the creative features and higher probability of adoption of new technology and functions by the largest cities.\\

The under-estimation of growth in the western part of the territory might be due to its integration within a larger area (the Eastern Europe) during the periods under study: our hypothesis here is that the centrality of western (post-)Soviet cities would then be minored in our model because it does not take into account the interactions with east-european cities (Warsaw, Prague, Bratislava, etc.) which formed altogether an economic system (even though the integration was always stronger within the FSU).

 \begin{table}[H]
\centering
\caption{Observed and Simulated populations of urban outliers in 1989}
\label{tab:res1959}
Positive Residuals\\
\begin{tabular}{|c|c|c|}
\hline
City &  Observed & Simulated \\ \hline
\small Naberezhnye Tchelny &  \small 500 000 &  \small 30 000  \\
\small Volgodonsk & \small 191 000 &  \small 36 000  \\
\small Chajkovskij &\small 86 000 & \small  19 000  \\
\small Toljatti& \small 685 000 &  \small 158 000  \\
\small Bratsk & \small 285 000 & \small  73 000  \\
\small Balakovo &  \small 197 000 &  \small 52 000  \\
\small Tihvin & \small 71 000 &  \small 20 000  \\
\small Chervonograd &\small 72 000 & \small  21 000  \\
\small Obninsk& \small 111 000 &  \small 32 000  \\
\small Staryjoskol & \small 174 000 & \small  53 000  \\ \hline
\end{tabular}.\\

Negative Residuals\\
\begin{tabular}{|c|c|c|}
\hline
City &Observed & Simulated \\ \hline
\small Zaozernyj & \small 16 000 & \small  54 000  \\
\small Gremjachnsk & \small 21 000 &  \small 56 000  \\
\small Atakent / Ilitch  & \small 15 000 &  \small 38 000  \\
\small Kizel & \small 37 000 & \small  88 000  \\
\small Cheremhovo  & \small 74 000 &  \small 172 000  \\
\small Ilanskij &  \small 18 000 &  \small 42 000  \\
\small Gornoaltajsk & \small 46 000 &  \small 102 000  \\
\small Volchansk &\small 15 000 & \small  32 000  \\
\small Zujevka& \small 16 000 &  \small 35 000  \\
\small Taldykorgai & \small 138 000 & \small  296 000  \\ \hline
\end{tabular}
\end{table}

 \begin{table}[H]
\centering
\caption{Observed and Simulated populations of urban outliers in 2010}
\label{tab:res1989}
Positive Residuals\\
\begin{tabular}{|c|c|c|}
\hline
City &  Observed & Simulated \\ \hline
\small Mirnyja &  \small 41 000 &  \small 12 000  \\
\small Sertolovo & \small 48 000 &  \small 16 000  \\
\small Beineu &\small 32 000 & \small  11 000  \\
\small Govurdak & \small 76 000 &  \small 28 000  \\
\small Serdar / Gyzylarbat & \small 98 000 & \small  37 000  \\
\small Bayramaly &  \small 131 000 &  \small 53 000  \\
\small Sarov & \small 92 000 &  \small 39 000  \\
\small Turkmenabat / Tchardjou &\small 427 000 & \small  185 000  \\
\small Astana / tselinograd & \small 613 000 &  \small 278 000  \\
\small Dashougouz& \small 275 000 & \small  126 000  \\ \hline
\end{tabular}.\\

Negative Residuals\\
\begin{tabular}{|c|c|c|}
\hline
City &Observed & Simulated \\ \hline
\small Sovetabad & \small 11 000 & \small  33 000  \\
\small Zhanatas & \small 21 000 &  \small 50 000  \\
\small Krasnozavodsk  & \small 13 000 &  \small 31 000  \\
\small Gagra & \small 11 000 & \small  25 000  \\
\small Nevelsk  & \small 12 000 &  \small 26 000  \\ 
\small Arkalyk  & \small 28 000 &  \small 59 000  \\ 
\small Chyatura  & \small 14 000 &  \small 28 000  \\ 
\small Aleksandrovsk Sahalinsk  & \small 11 000 &  \small 21 000  \\ 
\small Uglegorsk  & \small 10 000 &  \small 20 000  \\
\small Baikonyr  & \small 36 000 &  \small 67 000  \\ \hline
\end{tabular}
\end{table}

Finally, some individual cities appear as clear outlyers of the model, and could correspond to a profile that is too specific to be modelled by any generic mechanism.  For the first period, (cf. table \ref{tab:res1959}), the most obvious examples of singular trajectories are the cities which grew much faster than what was expected from their site, situation or interaction attributes. Indeed, Naberezhnye Tchelny, Volgodonsk, Toljatti or Bratsk owe their sudden development to political decisions to implement flagship projects: automobile industry mega-plants in Naberezhnye Tchelny (trucks) and Toljatti (cars), energy production sites in Volgodonsk (atomic power) and Bratsk (hydroelectric power station). These economic policies of the 1950s and 1960s led those cities to be four times as populated 30 years lates than what was expected from their interactions, resource or regional characteristics.\\

  For the second period, (cf. table \ref{tab:res1989}), Astana is a good example of a similar singular trajectory that we would not aim to simulate with generic urbanisation mechanisms, as it owes its booming growth to the decision of the Kazakh newly independent State to locate its headquarters in this city more central to the country (compared to Almaty). On the contrary, Baikonyr, also in Kazakhstan, has suffered from the cuts in the space industry (non-predictable at the urban level of our mechanisms). Other shrinking cities like Aleksandrovsk-Sahalinsk, Krasnozavodsk or Uglegorsk would require more detailed mechanisms of demographics (lack of birth and emigration) and economic cycles to be simulated adequately.\\

To summarise, there are particular types of urban trajectories that are not simulated well by the model because of its simplicity, and trajectories that are too specific to be modelled. We find that the exploration of our models, their calibration and the analysis of residuals has helped to identify those cities and to suggest some missing mechanisms.\\


\section{Conclusion}
\label{sec:conclu} 

\begin{quote}
\small
"Despite the fact that the experience of individual cities has become more varied internationally (at least within what might be called the mature economies) there is stronger evidence of a predictable pattern of change, determined by common causal factors, than might be expected given the diversity and variety of cities." \cite[p. 1342]{cheshire1999}
\end{quote}

Systems of cities have attracted a lot of attention from social modellers because of the regularity of their patterns. Instead of regarding the profusion of competing theories as a source of confusion (or suspicion), we proposed a framework to integrate complementary accounts of urbanisation processes into a modular agent-based model. This work of synthesis and testing within a virtual laboratory has been made possible by its automation and the extensive use of computation resources to calibration sixty-four models structures with empirical data on almost two thousands cities in the Former Soviet Union over the last fifty years. The model provides a basis for comparison of the different theories that can be augmented by new or alternative implementations of mechanisms. It could also be applied to different systems of cities (in space or time).\\

In the present study, we showed that the multi-modelling approach has helped identify and order the mechanisms that most probably generated the urban pattern or Soviet and post-Soviet urbanisation: situation effects of urban transition before 1991 and resource extraction afterwards. The intuitive mechanism of redistribution however has proved unsignificant. This method was finally useful to spot the most singular trajectories of cities that we interpret with empirical and monographic knowledge to refine the model.\\

To conclude, modelling experiments perform a radical compression of reality and an extreme simplification of individual trajectories, events and persons into synthetic aggregates. The ontological adequacy of the model to real life is therefore necessarily evaluated at an aggregated level that creates the problem of equifinality. No simulation model can elude this question, but everything should be tried to reduce its impact on what can be learned from the modelling experience.\\


\begin{center}
\bf \sc{Acknowledgments}
\end{center}
\small This work has been funded by the ERC Advanced Grant Geodivercity, the University Paris 1 Panth{\'e}on-Sorbonne and the Institut des Systèmes Complexes Paris Île-de-France. The results obtained in this paper were computed on the biomed and the \url{vo.complex-system.eu} virtual organization of the European Grid Infrastructure (\url{http://www.egi.eu}). We thank the European Grid Infrastructure and its supporting National Grid Initiatives (France-Grilles in particular) for providing the technical support and infrastructure.


\end{multicols}

\end{document}